\documentclass[aps,prc,fleqn,floatfix,twocolumn,superscriptaddress,twoside]{revtex4}

\usepackage[dvipdfm]{graphicx}
\usepackage{amsmath,amssymb,amsfonts}
\usepackage{color}

\newcommand{\mean}[1]{\left\langle #1 \right\rangle} 

\begin{document}

\title{Energy loss for heavy quarks in relation to light partons; is radiative energy loss for heavy quarks anomalous?}
\author{ Roy~A.~Lacey} 
\author{R.~Wei}
\author{ N.~N.~Ajitanand} 
\author{ J.~M.~Alexander}
\author{ X.~Gong}
\author{ J.~Jia}
\author{ A.~Mawi}
\author{ S.~Mohapatra}
\author{ D.~Reynolds}
\author{ S.~Salnikov}
\author{A.~Taranenko}

\affiliation{Department of Chemistry, 
Stony Brook University, \\
Stony Brook, NY, 11794-3400, USA}

\date{\today}
\begin{abstract}

The scaling properties of jet suppression measurements are compared for non-photonic electrons ($e^{\pm}$) 
and neutral pions ($\pi^0$) in Au + Au collisions at $\sqrt{s_{NN}}=200$\,GeV. For a broad range of 
transverse momenta and collision centralities, the comparison is consistent with jet quenching 
dominated by radiative energy loss for both heavy and light partons. Less quenching is 
indicated for heavy quarks via $e^{\pm}$; this gives an independent estimate of the transport 
coefficient $\hat{q}$ that agrees with its magnitude obtained from quenching of 
light partons via $\pi^0$'s. 



\end{abstract}
\maketitle


Partonic energy loss provides an important tomographic probe of the hot and dense 
matter produced in high energy nuclear collisions \cite{Gyulassy:1993hr}. Such energy 
loss is manifested as a suppression of high-$p_T$ hadron yields (jet quenching) in central 
and midcentral A+A collisions, when compared to the binary-scaled yields from p+p 
collisions \cite{Adler:2003qi,Adams:2003kv}. 
There is mounting evidence that the primary mechanism for light-parton energy loss is 
medium-induced gluon bremsstrahlung in the quark-gluon plasma (QGP) produced in 
A+A collision \cite{Gyulassy:1993hr,Baier:2000mf,Kovner:2003zj,Gyulassy:2000er,
Dokshitzer:2001zm, Wang:2001ifa, Majumder:2007hx,Lacey:2009ps}.
Therefore, measurements of jet quenching can provide important constraints 
for determining the transport coefficients of the QGP.

	Further insight into properties of the QGP can be gained from the production 
and propagation of particles carrying heavy quarks (charm or bottom). However, 
a central issue germane to the development of this probe, is the unsettled question of 
the quantitative difference between the quenching for light and heavy partons 
in hot QCD matter. Dokshitzer and Kharzeev \cite{Dokshitzer:2001zm} have argued that 
much less quenching is to be expected from heavy quarks because their associated 
gluon radiation is suppressed by the dead cone effect. 
In contrast, Zakharov \cite{Zakharov:2007pj,Aurenche:2009dj} argues that the radiative
energy loss for a finite-size QGP could even have an anomalous mass dependence
(ie. energy loss for heavy quarks greater than that for light 
quarks) which stem from quantum final state effects when the gluon formation time is 
comparable to the ``size'' of the QGP.

	Here, we use the published jet quenching measurements for non-photonic 
electrons ($e^{\pm}$) \cite{Adare:2008qa} and neutral pions ($\pi^0$) \cite{Adare:2006nq,Adare:2008cx}
at $\sqrt{s_{NN}}=200$\,GeV to compare, as well as to quantify the difference in the quenching patterns 
of the transverse momentum spectra for jets produced from scattered light- and heavy 
partons. 
We then use this difference to constrain an estimate of the transport coefficient 
$\hat{q}$ and the ratio of viscosity to entropy density $\eta/s$. We find that 
the value of these new estimates (of $\hat{q}$ and $\eta/s$) are similar to the ones  
obtained via jet quenching measurements for light partons via $\pi^0$'s \cite{Lacey:2009ps}.

	The nuclear modification factor ($R_{\rm AA}$) is used to quantify the magnitude of 
jet suppression in A+A collisions \cite{Adler:2003qi};
\[
   R_{\rm AA}(p_T) = \frac{1/{N_{\rm evt}} dN/dydp_{\rm T}}{\mean{T_{\rm AA}} d\sigma_{pp}/dydp_{\rm T}}, 
\]
where $\sigma_{pp}$ is the particle production cross section in p+p collisions 
and $\mean{T_{\rm AA}}$ is the nuclear thickness function
averaged over the impact parameter range associated with a given 
centrality selection
\[
\langle T_{AA}\rangle\equiv
\frac {\int T_{AA}(\mathbf{b})\, d\mathbf{b} }{\int (1- e^{-\sigma_{pp}^{inel}\, 
T_{AA}(\mathbf{b})})\, d\mathbf{b}}.
\]
The corresponding average number of nucleon-nucleon collisions, 
$\langle N_{coll}\rangle=\sigma_{pp}^{inel} \langle T_{AA}\rangle$,
is obtained with a Monte-Carlo Glauber-based 
model calculation \cite{Miller:2007ri,Alver:2006wh}.

	To facilitate our comparisons of the quenching patterns of the 
transverse momentum spectra for jets produced from scattered light ($l$)- and heavy ($h$)
partons, we exploit the formalism of Dokshitzer and Kharzeev (DK) \cite{Dokshitzer:2001zm};
\begin{eqnarray}
{R^h_{\rm AA}(p_T,L) \simeq}   \qquad {  } \nonumber \\ 
\exp \left[ - \frac{2 \alpha_s C_F}{\sqrt{\pi}} 
L\,\sqrt{\hat{q}\frac{{\cal{L}}_h}{p_T}} 
\,+ \frac{16 \alpha_s C_F}{9 \sqrt{3}}\, L \left( \frac{\hat{q}\> M^2}{M^2+p_T^2}\right)^{1/3} \right],
\label{eq:DK1}
\end{eqnarray}
where $\alpha_s$ is the strong interaction coupling strength, $C_F$ is the color factor,
$L$ is the path length [of the medium] that the parton traverses, $M$ is 
the mass of the heavy parton and $\hat{q}$ is the transport coefficient 
of the medium. 

The first term in the exponent in Eq.~\ref{eq:DK1} represents the
quenching of the transverse momentum spectrum which is the same for
both light and heavy partons \cite{Dokshitzer:2001zm},
\begin{eqnarray}
R^l_{\rm AA}(p_T,L) \simeq \exp \left[- \frac{2 \alpha_s C_F}{\sqrt{\pi}}\ 
L\,\sqrt{\hat{q}\frac{{\cal{L}}_l}{p_T}}\, \right] \nonumber \\
{\cal{L}} \approx \frac{d}{d\ln p_T} 
\ln \left[ \frac{d \sigma_{pp}}{d p_{T}^2}( p_{T})\right], 
\label{eq:DK2}	
\end{eqnarray}
(modulo the difference of the ${\cal{L}}$ parameters 
determined by the $p_{T}$ distributions in the vacuum).  The second
term in the exponent is specific for heavy quarks and is 
a direct consequence of the fact that the mass of the heavy quark 
leads to a suppression of gluon radiation. Hence the prediction that the 
suppression for the heavy hadron $p_T$ distribution should be smaller than 
that for light hadrons. 

	The magnitude of the heavy-to-light suppression factor can also 
be estimated as \cite{Dokshitzer:2001zm};
\begin{equation}
\frac{R^h_{\rm AA}(p_T,L)}{R^l_{\rm AA}(p_T,L)} \>\simeq\> 
 \exp \left[ \frac{16 \alpha_s C_F}{9 \sqrt{3}} L
\left( \frac{ \hat{q}\> \> M^2}{M^2+p_\perp^2}\right)^{1/3}  \right],
\label{eq:DK3}
\end{equation}
which clearly depends on the size ($L$) of the QGP medium and the properties encoded in 
the value of its transport coefficient $\hat{q}$.

	An essential point reflected in Eqs.~\ref{eq:DK1} - \ref{eq:DK3} is that they give specific 
testable scaling predictions for the $p_T$ and $L$ dependence 
of $R^{h,l}_{\rm AA}(p_T,L)$, and the heavy-to-light suppression ratio \cite{Lacey:2009ps}. 
That is, $\ln\left[R^{h,l}_{\rm AA}(p_T,L)\right]$ should scale as $L$ and $1/\sqrt{p_T}$ 
respectively, and the heavy-to-light suppression ratio should scale as $L$. 
As discussed below, these scaling properties provide crucial validation tests of our 
analysis framework, as well as an important constraint for estimating $\hat{q}$ and $\eta/s$.

\begin{figure}[t]
\includegraphics[width=1.0\linewidth]{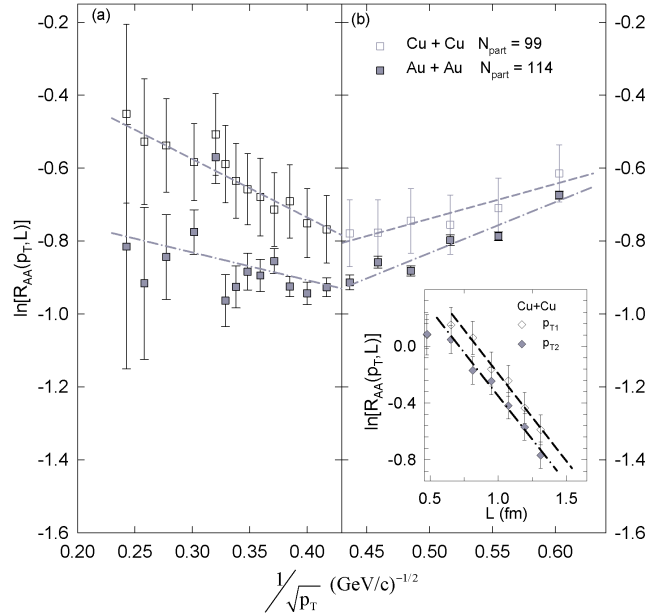}
\caption{$\ln\left[R_{\rm AA}(p_T,L)\right]$ vs. $1/\sqrt{p_T}$ for $p_T \agt 5$~GeV/c (a)
and $2.75 \alt p_T \alt 5$~GeV/c (b) for $\pi^0$'s produced in Cu+Cu (N$_{\rm part} = 99$) 
and Au+Au (N$_{\rm part} = 114$) collisions. 
The inset shows $\ln\left[R_{\rm AA}(p_T,L)\right]$ vs. $L$ for $\pi^0$'s 
with  $p_{T1}= 9.25$ GeV/c and $p_{T2}=5.75$ GeV/c in Cu+Cu collisions. 
Error bars are statistical only. 
The curves in each panel are linear fits to each data set (see text).
}
\label{Fig1}
\end{figure}

	The result of one such test for Cu+Cu and Au+Au collisions is given in Fig.~\ref{Fig1},
where we show a plot of $\ln\left[R_{\rm AA}(p_T,L)\right]$ vs. $1/\sqrt{p_T}$ for 
$\pi^0$'s. Here, results are shown for $p_T \agt 5$~GeV/c in panel (a) and 
for $2.75 \alt p_T \alt 5$~GeV/c 
in panel (b). Both panels indicate the predicted linear dependence on $1/\sqrt{p_T}$
(cf. Eq.~\ref{eq:DK2}), albeit with different slopes for each $p_T$ range. 
A comparison of the results for the two collision systems also indicate similar 
quenching magnitudes when the size of the respective collision medium is comparable.
Note that better agreement is achieved if the small difference in the length estimate  
arising from the respective centrality cut is taken into account.

	The inset in Fig.~\ref{Fig1}(b) confirms the predicted linear dependence 
of $\ln\left[R_{\rm AA}(p_T,L)\right]$ on $L$ for Cu+Cu collisions. A similar 
dependence has been reported earlier for Au+Au collisions \cite{Lacey:2009ps}.
Here, as in Ref. \cite{Lacey:2009ps}, we have used the transverse size of the 
system $\bar{R}$ as an estimate of the operative path length $L$. To obtain this 
estimate, the number of participant nucleons $N_{\rm part}$ was estimated for each 
centrality selection, via a Monte-Carlo Glauber-based model \cite{Miller:2007ri,Alver:2006wh}. 
The corresponding transverse size $\bar{R}$ was then determined from the distribution 
of these nucleons in the transverse ($x,y$) plane via the 
same Monte-Carlo Glauber model:
%
$
1/\bar{R} = \sqrt{\left(\frac{1}{\sigma_x^2}+\frac{1}{\sigma_y^2}\right)},
$
%
%
%
%
where $\sigma_x$ and $\sigma_y$ are the respective root-mean-square widths of
the density distributions. For these evaluations, the initial entropy profile in 
the transverse plane was assumed to be proportional to a linear combination
of the number density of participants and binary 
collisions \cite{Hirano:2009ah,Lacey:2009xx}. The latter assures that the entropy 
density weighting is constrained by multiplicity measurements.

\begin{figure}[t]
\includegraphics[width=1.0\linewidth]{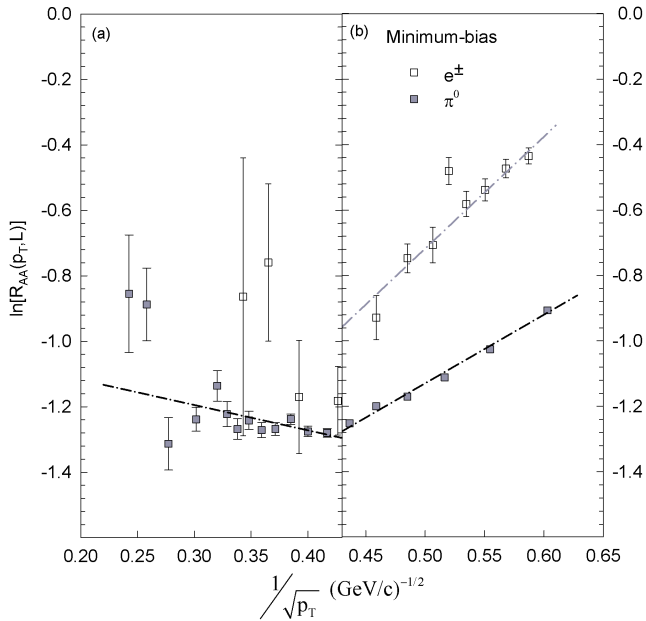}
\caption{$\ln\left[R_{\rm AA}(p_T,L)\right]$ vs. $1/\sqrt{p_T}$ for $p_T \agt 5$~GeV/c (a)
and $2.75 \alt p_T \alt 5$~GeV/c (b) for minimum-bias Au+Au 
collisions. Error bars are statistical only. 
The curves in each panel are linear fits to each data set (see text).} 
\label{Fig2}
\end{figure}

	 Figure \ref{Fig2} shows a comparison of the $R_{\rm AA}$ measurements obtained for 
neutral pions ($\pi^0$) and non-photonic electrons ($e^{\pm}$) in minimum-bias Au+Au 
collisions \cite{Adare:2008qa,Adare:2006nq}. As in Fig.~\ref{Fig1}, results are shown for 
$p_T \agt 5$~GeV/c in panel (a) and for $2.75 \alt p_T \alt 5$~GeV/c in panel (b).  
The latter $p_T$ range for $e^{\pm}$ is dominated by decay contributions from D-mesons. However, 
as demonstrated in recent measurements \cite{Adare:2009ic,Zhang:2008cb}, 
there is an increasing role of the B-meson contributions to the non-photonic elecron 
spectrum for $p_T \agt 5$~GeV/c. Such contributions could serve to complicate the 
interpretation of the $R_{\rm AA}$ measurements for $e^{\pm}$ above $p_T \sim 5$~GeV/c.

	The $\pi^0$ measurements in Fig.~\ref{Fig2}(a) show the predicted linear dependence 
on $1/\sqrt{p_T}$. However, the statistical significance of the data for $e^{\pm}$ does 
not allow a similarly decisive conclusion in this $p_T$ range. The measurements shown in Fig.~\ref{Fig2}(b)
contrast with those of Fig.~\ref{Fig2}(a). Here, both data sets confirm the 
the predicted linear dependence on $1/\sqrt{p_T}$. Even more striking is the observation
that the magnitude of the quenching for $\pi^0$ is significantly larger than that for 
$e^{\pm}$. This observation is in accord with the prediction of 
Dokshitzer and Kharzeev (cf. Eq.~\ref{eq:DK1} and Ref.~\cite{Dokshitzer:2001zm}) that 
the suppression of heavy hadrons should be less than that for light hadrons. 
As discussed below, it is also congruent with the observation that the magnitude of the 
azimuthal anisotropy for $e^{\pm}$ (characterized by the second Fourier coefficient $v_2$)  
is much less than that for pions (for $p_T \agt 2.5$~GeV/c) \cite{Adare:2008qa}.

\begin{figure}[!t]
\includegraphics[width=0.95\linewidth]{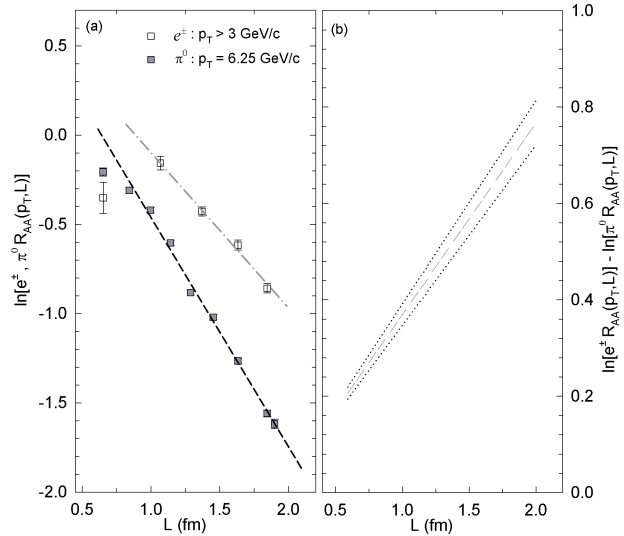}
\caption{(a) $\ln\left[R_{\rm AA}(p_T,L)\right]$ vs. $L$ for $\pi^0$ and 
non-photonic electrons ($e^{\pm}$) as indicated. The dashed and dot-dashed curves 
indicates a linear fit to each data set. The error bars in both panels are statistical only.
(b) $\ln\left[\left( \frac{R^h_{\rm AA}(p_T,L)}{R^l_{\rm AA}(p_T,L)}\right) \right]$ vs. $L$.
This ratio is obtained from the fits to the $\pi^0$ and $e^{\pm}$ 
data sets in panel (a). The dotted lines represent the statistical error band.  
} 
\label{Fig3}
\end{figure}

Figure \ref{Fig3} compares and contrasts the $L$ dependence of $\ln\left[R_{\rm AA}(p_T,L)\right]$ for 
$\pi^0$ and $e^{\pm}$ for the $p_T$ selections indicated. We emphasize here that 
kinematic studies show that the $p_T$ selection for $e^{\pm}$ (cf. Fig.~\ref{Fig3}) leads to  
a dominant $e^{\pm}$ decay contribution from D-mesons of higher $p_T$. 
The dashed and dot-dashed curves in panel (a) are linear fits to the data 
set for $\pi^0$ and $e^{\pm}$ respectively (the data point for the most peripheral collision 
is treated as an outlier). The dashed curve in panel (b) shows the $L$ dependence of the ratio 
$\ln\left[\left( \frac{R^h_{\rm AA}(p_T,L)}{R^l_{\rm AA}(p_T,L)}\right) \right]$ obtained 
from the fits in panel (a).
These curves validate the predicted linear dependence on 
$L$ (cf. Eqs. \ref{eq:DK1} - \ref{eq:DK3}), as well 
as the attendant difference in slope for $\pi^0$ and $e^{\pm}$. We interpret 
the latter as an independent indication that the mechanism for jet quenching is 
dominated by radiative energy loss.

		The fits to the data in Fig.~\ref{Fig3} (a) indicate intercepts 
of $L = 0.6 \pm 0.1$~fm and $L = 0.9 \pm 0.15$~fm for the light- and heavy 
mesons respectively. This suggests a minimum path length requirement 
for the initiation of jet suppression for both light and heavy partons, but with the 
possibility of a larger path length requirement for the heavy quark.  
The corresponding ratio $\ln\left[\left( \frac{R^h_{\rm AA}(p_T,L)}{R^l_{\rm AA}(p_T,L)}\right) \right]$, 
plotted as a function of $L$ in Fig.~\ref{Fig3} (b), yields a slope value of $0.4 \pm 0.04\,{\rm fm}^{-1}$.   
%
	This value reflects the difference in slope for $\pi^0$ and $e^{\pm}$ shown in Fig.~\ref{Fig3}(a).
It is noteworthy that the latter is fully compatible with the observed difference of about a factor of 
two  in  the measured magnitude of $v_2$ for $\pi^0$ and $e^{\pm}$ for $p_T \agt 2.5$~GeV/c  \cite{Adare:2008qa}. 
Thus, it provides further confirmation that the azimuthal anisotropy of particle yields (for the meson 
$p_T$ ranges of interest) stem from the path-length dependence of jet quenching. This observation further 
underscores the importance of reporting jet suppression measurements in conjunction with anisotropy ($v_2$)
measurements at high $p_T$.

	To obtain an ``independent'' estimate of the magnitude of $\hat{q}$, 
we assume a negligible $e^{\pm}$ contribution from B-mesons and use the slope ($0.4 \pm 0.04\,{\rm fm}^{-1}$) 
extracted from Fig.~\ref{Fig3}(b) in concert with Eq.~\ref{eq:DK3}. 
This gives the value $\hat{q} \approx 0.73 \pm 0.12$~GeV$^2$/fm
for the values $\alpha_s = 0.3$ \cite{Bass:2008rv}, 
$C_F = 4/3$, $M = 1.5$~GeV and $\left<p_T\right> \approx 6$~GeV/c (for D-mesons). 
This estimate of $\hat{q}$ is comparable to the recent estimate of $\approx 0.75$~GeV$^2$/fm 
obtained solely from $\pi^0$ suppression measurements with the same definition of $L$
\cite{Lacey:2009ps}. It is also compatible 
with the value of 1 - 2 GeV$^2$/fm obtained from fits to hadron suppression data within the 
framework of the twist expansion \cite{Zhang:2007ja,Majumder:2007ae}. 

	The value for $\hat{q}$ can be used to estimate the ratio of the shear viscosity ($\eta$) to 
the entropy density ($s$) as \cite{Majumder:2007zh};
\[
\frac{\eta}{s} \approx 1.25 \frac{T^3}{\hat{q}}
\]
where $T$ is the temperature. 
This gives the estimate $4\pi\frac{\eta}{s} \approx 1.4 \pm 0.4$ 
for $T \sim 220$~MeV \cite{Adare:2008fqa}, which is comparable to the 
value estimated via $\pi^0$ suppression measurements \cite{Lacey:2009ps} and  
flow measurements \cite{Lacey:2009xx,Lacey:2006bc}. We conclude therefore, that the 
relatively short mean free path in the QGP \cite{Lacey:2009xx} leads to 
hydrodynamic-like flow with small shear viscosity, as well as significant 
quenching of both light and heavy partons.

	In summary, we have compared the scaling properties of jet quenching 
for both light- and heavy partons, via suppression measurements for $\pi^0$ and $e^{\pm}$. 
Our comparisons confirm the predicted $p_T$ and $L$ dependence for medium-induced gluon 
radiation in a hot and dense QGP. The difference in the magnitude of the quenching for 
light- and heavy partons, do not give a strong indication for anomalous heavy quark energy loss
in the $p_T$ range studied. Instead, it is compatible with the prediction that gluon radiation 
from heavy quarks is suppressed by the dead cone effect. This difference also gives an 
estimate $\hat{q} \sim 1$~GeV$^2$/fm, which is essentially the same as that obtained 
via $\pi^0$ suppression measurements. 
Future detailed measurements of D- and B-mesons at high $p_T$, as well as model calculations 
which take full account of the reaction dynamics, will undoubtedly provide more detailed 
mechanistic insights for heavy quark energy loss and improved constraints for the 
transport properties of the QGP.

\section*{Acknowledgments}
We thank D.E. Kharzeev for helpful discussions.
This work was supported by the US DOE under contract DE-FG02-87ER40331.A008.
 

\bibliography{QhatRefs}
\end{document}